\pdfoutput=1
\documentclass[11pt]{article}

\usepackage[preprint]{acl}

\usepackage{times}
\usepackage{latexsym}

\usepackage[T1]{fontenc}
\usepackage[utf8]{inputenc}
\usepackage{multicol}
\usepackage{multirow}
\usepackage{microtype}

\usepackage{inconsolata}

\usepackage{graphicx}

\title{Beyond Turing Test: Can GPT-4 Sway Experts' Decisions?}

\author{
 \textbf{Takehiro Takayanagi,\textsuperscript{1,2}}
 \textbf{Hiroya Takamura,\textsuperscript{2}}
 \textbf{Kiyoshi Izumi,\textsuperscript{1}}
 \textbf{Chung-Chi Chen\textsuperscript{2}}
\\
\\
 \textsuperscript{1}The University of Tokyo, \\
 \textsuperscript{2}National Institute of Advanced Industrial Science and Technology
\\
 \texttt{takayanagi-takehiro590@g.ecc.u-tokyo.ac.jp, takamura.hiroya@aist.go.jp,}\\ \texttt{izumi@sys.t.u-tokyo.ac.jp, c.c.chen@acm.org}\\
}

\begin{document}
\maketitle
\begin{abstract}
In the post-Turing era, evaluating large language models (LLMs) involves assessing generated text based on readers' reactions rather than merely its indistinguishability from human-produced content. This paper explores how LLM-generated text impacts readers' decisions, focusing on both amateur and expert audiences. Our findings indicate that GPT-4 can generate persuasive analyses affecting the decisions of both amateurs and professionals. Furthermore, we evaluate the generated text from the aspects of grammar, convincingness, logical coherence, and usefulness. The results highlight a high correlation between real-world evaluation through audience reactions and the current multi-dimensional evaluators commonly used for generative models. Overall, this paper shows the potential and risk of using generated text to sway human decisions and also points out a new direction for evaluating generated text, i.e., leveraging the reactions and decisions of readers. We release our dataset to assist future research. 

\end{abstract}

\section{Introduction}
\label{sec:intro}
Large language models (LLMs) have demonstrated impressive performance, and the Turing test has become less reliable for evaluating LLM-generated text~\cite{tikhonov-yamshchikov-2023-post}. In other words, pursuing the generation of content indistinguishable from that produced by humans is no longer the goal in the post-Turing era. Nowadays, we should evaluate LLM-generated text using the same criteria applied to human-generated text. 
In the real world, these criteria are always related to readers' reactions. For example, the number of views is an important evaluation metric for YouTube videos, the number of likes is the evaluation metric for social media editors, and the obtained donations are the best metrics for crowdfunding proposals. Following this line of thought, this paper provides a pilot exploration of linking generated text with readers' reactions.
Going a step further, the behaviors and reactions of common people and experts are very different~\cite{snow-etal-2008-cheap,aguda-etal-2024-large-language}. To analyze this difference, we include the reactions of both amateurs and experts for in-depth discussions.

\begin{figure}[t]
  \begin{center}
    \includegraphics[clip,keepaspectratio,width=\linewidth]{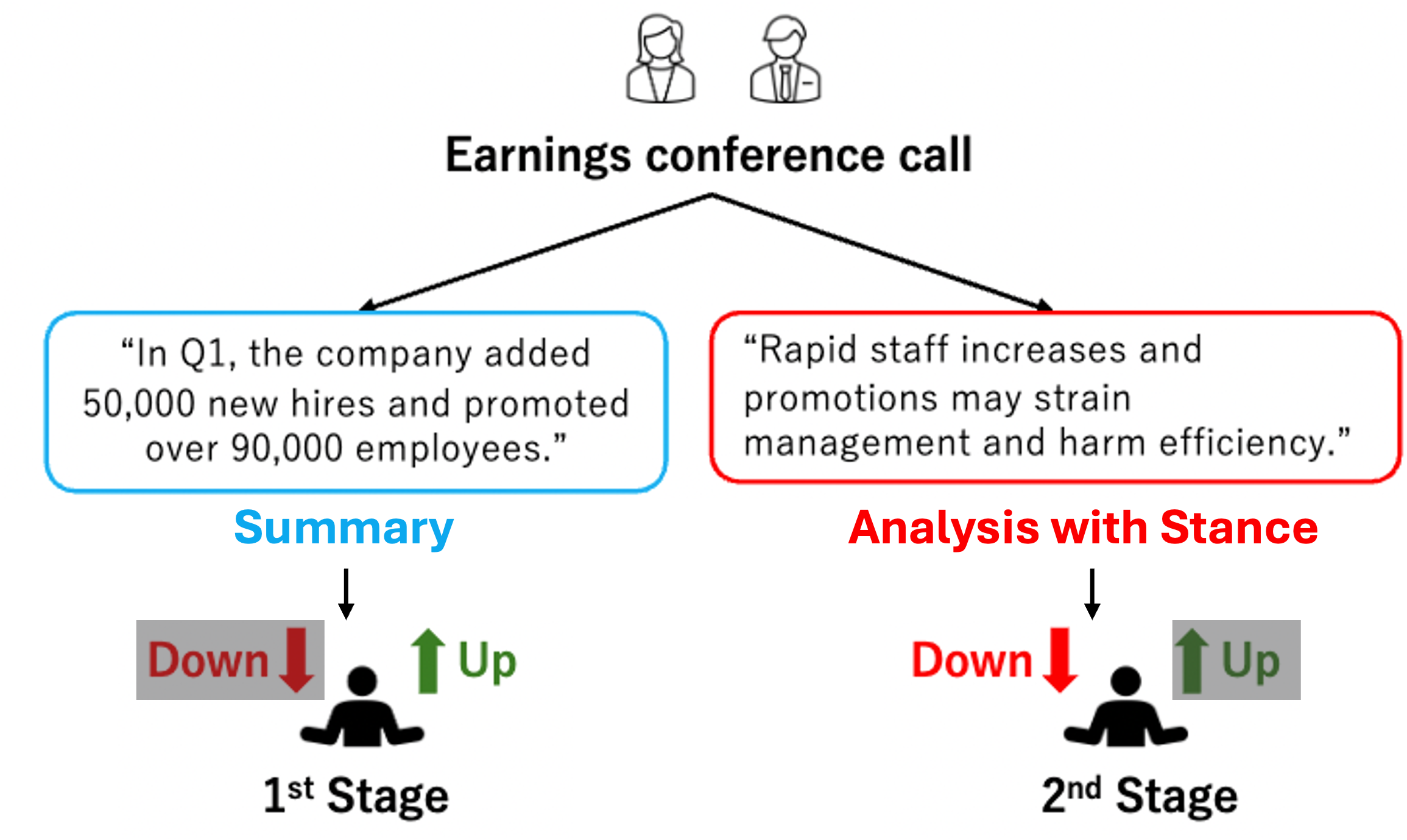}
    \caption{Design of experiments.}
    \label{fig:Design of experiments}
  \end{center}
\end{figure}

Inspired by previous studies~\cite{kimbrough2005effect,keith-stent-2019-modeling}, earnings conference calls (ECCs)—meetings among company managers and professional analysts to discuss the latest operations and future plans—affect both amateur and professional investors' decisions. This scenario fits our scope, which aims to discuss how the information provided influences amateurs' and experts' decisions. Therefore, we designed our experiments based on ECCs. 
Figure~\ref{fig:Design of experiments} illustrates the design of the experiment. We first provide an objective summary of the ECC and ask investors to predict whether to increase or decrease based on the given summary. Then, we provide a subjective analysis for the same ECC to investors and ask them to decide whether they want to change their decisions. Our results reveal that GPT-4~\cite{openai2023gpt4} can generate persuasive analysis that sways both amateurs' and professionals' decisions.

Given that many recent studies~\cite{zhong-etal-2022-towards,chan2023chateval} propose evaluating generated text by scoring, we also assess the generated text from both objective (grammar) and subjective (convincingness, logical coherence, and usefulness) aspects. Our results indicate that both objective and subjective evaluation metrics do highly correlate with the decisions. 
The high correlation between multi-dimensional evaluators
 and real-world evaluations (audience/reader reactions) in our experiment highlights the potential of using readers’ reactions as an evaluation method.
% This reveals the gap between real-world evaluation (audience/reader reaction) and the current multi-dimensional evaluators in most studies for generative models. 

To sum up, this paper focuses on the following research questions:

\vspace{2mm}

\noindent \textbf{(RQ1)}: To what extent does state-of-the-art LLM-generated text sway people's decisions?

\vspace{2mm}

\noindent \textbf{(RQ2)}: Are the generated text's influences on amateurs and professionals different? 

\vspace{2mm}

\noindent \textbf{(RQ3)}: Does the recent popular evaluation approach align with reactions?

\section{Related Work}
The impact of text information on financial markets is a widely studied topic. Research has shown that different kinds of text data, from social media to financial news, can affect both trading algorithms and investor behavior~\citep{karppi2016social, fakenews}. Furthermore, the effects of bullish articles, created as part of stock promotion schemes, have been examined for their ability to draw investor attention and influence the market~\citep{clarke2020fake}.
The relationship between artificial intelligence and investor decision-making is another key area of research. \citet{lai2023} reviews recent studies exploring how AI and humans interact in various domains, including finance. Additionally, research examines how machine learning results affect investor choices~\citep{biran2017human}.
Despite considerable research into NLP applications in finance, the influence of text on financial markets, and the interaction between AI and investors, there remains a gap in studies specifically examining the impact of LLMs on investors' decisions. Our paper addresses this gap by proposing a novel evaluation framework.

\section{Experimental Design}

\subsection{Dataset}
We adopt the ECTSum~\cite{mukherjee-etal-2022-ectsum} dataset as the base for our experiment. In ECTSum, there are 2,425 ECC transcripts with professional journalist-written summaries. We manually aligned these data with the professional analysis reports on the Bloomberg Terminal,\footnote{https://www.bloomberg.com/professional/products/bloomberg-terminal/} which is one of the largest financial information vendor platforms. Finally, we obtained 234 instances containing the corresponding analysis reports. GPT-4~\cite{openai2023gpt4}~\footnote{We utilize \texttt{gpt-4-1106} in our experiments.} was used to generate the analysis by providing the ECC transcript and the stance (Overweight/Underweight), where overweight (underweight) denotes the suggesting increasing (decreasing) stock prices. Inspired by \citet{kogan2023social}, providing analysis from a certain aspect is rational, but intent to promote the analysis from a certain aspect is illegal. Thus, in addition to having GPT-4 act as a professional analyst, we also had GPT-4 act as a promoter to render and write an analysis with a stronger stance.\footnote{All prompts are available in Appendix~\ref{appendix:prompts}.}

\subsection{Evaluation Paradigm}
We recruited five financial experts with over five years of industry experience and eight students with academic backgrounds in finance for the experiment. There are two stages in each round of the experiment. In the first stage, participants are presented with neutral summaries, either professional journalist-written or GPT-4-generated summaries. Participants are asked to decide whether to increase or decrease the stock of the company within three-day trading period following the conference date. In the second stage, participants received a document with an investment stance pertaining to the same ECC as in the first stage. 
The documents are either professional analysis reports or GPT-4-generated analyses with stance. 
They were again asked to make a decision for the same three-day period. Here, a three-day setting was selected based on the empirical study of previous work~\cite{birru2022analyst}, which supports that the market reflects information within three days. In this way, we can answer (RQ1) and (RQ2) by analyzing the change between the two stages and the difference between students (amateurs) and experts (professionals). 

The basic salary of participants is 180\% of the minimum salary stipulated by law. To mimic real-world incentives and motivate participants to try their best to make the decision, their salary will increase to 270\% of the minimum salary stipulated by law as a reward if they make the correct decisions for 50\% of instances. To ensure the fairness of the experiment, we anonymized the stocks in all documents. This is intended to prevent participants from applying external knowledge, ensuring that their decisions are based solely on the information provided within the documents.

% \begin{table}[t]
%   \centering
%   \resizebox{\columnwidth}{!}{
%     \begin{tabular}{l|rrrr}
%     2nd Stage Source      & \multicolumn{1}{c}{All} & \multicolumn{1}{c}{Amateur} & \multicolumn{1}{c}{Expert} & \multicolumn{1}{c}{Veteran} \\
%     \hline
%     GPT-4 & 22.2\% & 25.7\% & 16.7\% & 8.7\% \\
%     Analyst & 35.9\% & 33.3\% & 40.0\% & 33.3\% \\
%     \end{tabular}%
%     }
%   \caption{Ratio of changing decisions in the second stage.}
%   \label{tab:changing}%
% \end{table}%

% Revised table
\begin{table}[t]
  \centering
  \resizebox{\columnwidth}{!}{
    \begin{tabular}{l|rrrr}
    2nd Stage Source      & \multicolumn{1}{c}{All} & \multicolumn{1}{c}{Amateur} & \multicolumn{1}{c}{Expert} & \multicolumn{1}{c}{Veteran} \\
    \hline
    GPT-4 & 28.7\% & 31.3\% & 24.7\% & 15.6\% \\
    Analyst & 26.3\% & 25.0\% & 28.3\% & 21.2\% \\
    \end{tabular}%
    }
  \caption{Ratio of changing decisions in the second stage.}
  \label{tab:changing}%
\end{table}%

\begin{table}[t]
  \centering
  \small
    \begin{tabular}{l|rrr}
    \multicolumn{1}{c|}{Change} & \multicolumn{1}{c}{Amateur} & \multicolumn{1}{c}{Expert} & \multicolumn{1}{c}{Veteran} \\
    \hline
    Upward & 24.1\% & 42.3\% & 44.4\% \\
    Downward & 75.9\% & 57.7\% & 55.6\% \\
    \end{tabular}%
  \caption{Direction of the change.}
  \label{tab:updown}%
\end{table}%

\begin{table}[t]
  \centering
  \resizebox{\columnwidth}{!}{
    \begin{tabular}{ll|rrrr}
    \multicolumn{1}{c}{Prompt} & \multicolumn{1}{c|}{Stance} & \multicolumn{1}{c}{All} & \multicolumn{1}{c}{Amateur} & \multicolumn{1}{c}{Expert} & \multicolumn{1}{c}{Veteran} \\
    \hline
    \multirow{2}[2]{*}{Analysis} & Overweight & 12.5\% & 11.8\% & 13.6\% & 6.6\% \\
          & Underweight & 37.1\% & 50.0\% & 16.7\% & 7.6\% \\
    \hline
    \multirow{2}[1]{*}{Promote} & Overweight & 23.7\% & 18.9\% & 31.8\% & 26.7\% \\
          & Underweight & 40.4\% & 42.9\% & 36.4\% & 21.4\% \\
    \end{tabular}%
    }
  \caption{Influence of prompts and stances.}
  \label{tab:prompts}%
\end{table}%

\begin{table}[t]
  \centering
  \small
    \begin{tabular}{l|rrr}
    Stage & \multicolumn{1}{c}{Amateur} & \multicolumn{1}{c}{Expert} & \multicolumn{1}{c}{Veteran} \\
    \hline
    1st   & 61.2\% & 61.3\% & 62.2\% \\
    2nd   & 45.8\% & 44.7\% & 51.1\% \\
    \end{tabular}%
    \caption{Accuracy of decisions.}
  \label{tab:Accuracy}%
\end{table}%

\section{Behavioral Experiment}
\subsection{Preprocessing}
\label{sec:Preprocessing}
The estimated cost of conducting experiments for all 234 instances is approximately 4,000 USD, which is prohibitively expensive. Therefore, we first adopt the Hierarchical Transformer-based Multi-task Learning model (HTML), utilized in financial forecasting based on ECCs~\citep{yang2020html}, to simulate the experiment. To simulate the first stage of the experiment, we use additional neutral summaries from the ECTSum dataset to train the model. During the testing phase, we use the neutral summary and the analysis with stance as input to simulate the second stage. If the model's decision changes when given a summary and analysis, we select this summary-analysis pair for the human behavioral experiment. Ultimately, we have 75 instances for the experiment, reducing the cost to about 1,280 USD.\footnote{More details about the setting of HTML are shown in Appendix \ref{sec:html}.}

\begin{table}[t]
  \centering
  \resizebox{\columnwidth}{!}{
    \begin{tabular}{ll|rrrr}
    \multicolumn{1}{c}{Annotator} & \multicolumn{1}{c|}{Source} & \multicolumn{1}{c}{Grammatical} & \multicolumn{1}{c}{Convincing} & \multicolumn{1}{c}{Logical} & \multicolumn{1}{c}{Useful} \\
    \hline
    \multirow{3}[2]{*}{Amateur} & Analysis (GPT-4) & 4.44  & 4.13  & 4.02  & 4.06 \\
          & Promote (GPT-4) & 4.47  & 4.23  & 4.16  & 4.20 \\
          & Analyst & 3.92  & 3.22  & 3.30  & 3.43 \\
    \hline
    \multirow{3}[2]{*}{Expert} & Analysis (GPT-4) & 3.65  & 2.80  & 3.04  & 2.84 \\
          & Promote (GPT-4) & 3.79  & 2.95  & 3.22  & 3.06 \\
          & Analyst  & 3.78  & 3.48  & 3.61  & 3.65 \\
    \hline
    \multirow{3}[1]{*}{Veteran} & Analysis (GPT-4) & 3.71  & 2.78  & 3.03  & 2.46 \\
          & Promote (GPT-4) & 3.79  & 2.95  & 3.22  & 3.06 \\
          & Analyst & 4.06  & 3.93  & 4.09  & 3.97 \\
    \end{tabular}%
    }
    \caption{Multi-dimensional evaluation.}
  \label{tab:multi-dimensional}%
\end{table}%

\subsection{Results and Analysis}
\label{sec:result}

Table~\ref{tab:changing} provides answers to RQ1 and RQ2. All experts have worked in the financial industry for more than five years, and we further group three experts with over ten years of experience as Veterans. 
First, the analysis written by professional analysts has a higher chance of changing experts' decisions.
Second, amateurs are more likely to change their decisions based on GPT-4-generated analysis. Additionally, more experienced investors are less influenced by GPT-4-generated analysis. These results indicate that GPT-4's analysis may suffice for amateur scenarios but is still far from professional standards. It also echoes previous studies' concerns about human evaluation quality in natural language generation research~\cite{snow-etal-2008-cheap,howcroft-etal-2020-twenty}, as many studies still evaluate models' outputs on crowdsourcing platforms. In other words, our results suggest that the analysis impacting amateurs may not be the focus for experts. 

Table~\ref{tab:updown} further shows the direction of their decision changes. Upward (Downward) denotes a change in their predictions from degrease (increase) to increase (decrease). Overall, investors are more sensitive to underweight analysis, i.e., information that may negatively impact the company. However, the ratio between amateurs and experts is significantly different. This indicates that amateurs are very sensitive to negative information. This raises a potential risk of using LLMs to generate analysis for the general public. The generated underweight analysis has a higher potential to sway amateur investors' decisions, and our results provide evidence supporting the U.S. Department of Treasury's concerns about the risks of artificial intelligence in the financial services sector.\footnote{https://home.treasury.gov/news/press-releases/jy2393} Imagining that automatically generated underweight analyses are widely distributed on online platforms, it may lead to higher market volatility and harm market stability. 

To conduct an in-depth analysis of the risk, we further use GPT-4 to write promoting reports for the given stance. Table~\ref{tab:prompts} shows the comparison. First, underweight analysis influences investors much more than overweight analysis. Second, analysis with a strong tone sways experts' decisions more than pure analysis, regardless of the given stance. This reveals the potential of LLMs in influencing professionals' decisions. 

Finally, as mentioned in Section~\ref{sec:Preprocessing}, we only focus on the pairs that lead the model to change decisions in spite of the accuracy. Thus, the analysis given in the second stage is not selected to lead investors to make wrong decisions. In Table~\ref{tab:Accuracy}, we show the accuracy of their decisions. The results reveal that investors make accurate trading decisions based on neutral summaries, and the analysis with stances may hurt the accuracy of their decisions. Based on this result, we want to highlight the risk of using generated analysis for financial decisions.

\subsection{Generated Text Evaluation}
Recently, many studies have scored generated text from multiple aspects~\cite{zhong-etal-2022-towards,chan2023chateval} to evaluate the quality of the generated documents. To answer (RQ3), we asked participants to annotate the given analysis from four aspects: grammar, convincingness, logical coherence, and usefulness. The score ranges from 1 to 5 (Discrete), with higher scores indicating better quality. Table~\ref{tab:multi-dimensional} shows the average scores of different groups of participants for different sources.

First, from the objective aspect, i.e., grammar, GPT-4 achieves a level similar to that of professional analysts, regardless of the group of annotators. However, from the subjective aspects, amateurs and experts have different opinions on GPT-4-generated and analyst-written analyses. Amateurs provide higher scores for GPT-4-generated text, while experts provide higher scores for analyst-written analysis. These results highlight the difference between amateurs and experts. Given this evidence, future works should reconsider the design of the human annotation process.

Second, compared with the results in Section~\ref{sec:result}, experts change their decisions more frequently when analysts' reports are provided in the second stage, and these reports are considered more convincing, logical, and useful. The situation is similar for amateurs; GPT-4-generated analysis gets higher scores and leads to more changes in amateurs' decisions. 
This indicates that scores and reactions are correlated in our experiment. 
% This shows the potential of using readers' reactions as an evaluation method for forward-looking analysis. 
The correlation between scores and reactions in our experiment highlights the potential of using these reactions to evaluate forward-looking analyses, including predicting future stock trajectories with rationales.
Finally, the experts' multi-dimensional evaluation scores also show the gap between state-of-the-art LLMs and professional analysts in writing analysis.

To check the agreement, each pair was annotated by at least two experts and two amateurs. We calculated Krippendorff's Alpha~\citep{krippendorff2011computing}, and the results are shown in Table~\ref{tab:Agreement}. The agreement on grammatical scores is very high regardless of the annotators. This suggests that evaluating generated text from objective aspects is effective, as most studies did before the LLM era. However, the agreement on subjective metrics is quite low, even among experts. This indicates the problem of conducting human evaluation from subjective aspects, as different people have different opinions. Following the discussion of \citet{amidei-etal-2018-rethinking}, the low agreement for complex generated text does not imply it is an insufficient evaluation metric, but it is natural after the generated text passes the Turing test. We hope the discussion in this paper can open different perspectives on generated text evaluation, particularly using readers' reactions as evaluation metrics.

\begin{table}[t]
  \centering
    \resizebox{\columnwidth}{!}{
    \begin{tabular}{l|rrrr}
          & \multicolumn{1}{c}{Grammatical} & \multicolumn{1}{c}{Convincing} & \multicolumn{1}{c}{Logical} & \multicolumn{1}{c}{Useful} \\
    \hline
    All   & 0.654 & 0.262 & 0.262 & 0.237 \\
    \hline
    Amateur & 0.505 & 0.109 & 0.136 & 0.179 \\
    Expert & 0.769 & 0.317 & 0.391 & 0.169 \\
    Veteran & 0.754 & 0.118 & 0.126 & 0.027 \\
    \end{tabular}%
    }
    \caption{Agreement among annotators.}
  \label{tab:Agreement}%
\end{table}%

\section{Conclusion}
This paper advocates for a nuanced approach to evaluating LLM-generated text and emphasizes the importance of real-world reactions as well as traditional evaluative metrics. By understanding and addressing the differences in how amateurs and experts perceive and are influenced by LLM-generated content, we can better harness the capabilities of these models while safeguarding against their potential pitfalls. Future research should continue exploring these dynamics, particularly focusing on the ethical implications and regulatory frameworks necessary to guide the responsible use of LLMs in decision-critical applications.

\section*{Limitations}

First, the scope of our study is restricted to ECCs within the financial sector. Although this context is highly relevant for examining decision-making processes, the results may not be directly transferable to other domains where different types of information and decision-making criteria are involved. Future studies should explore a broader range of contexts to validate and expand upon our findings.
Second, the sample size for our human behavioral experiment, though carefully selected to balance cost and representativeness, remains limited with 75 instances. This constraint may affect the statistical power and precision of our conclusions. Larger-scale studies are needed to confirm the trends and patterns observed in our research.
Third, the evaluation of generated text involved subjective metrics such as convincingness, logical coherence, and usefulness, which inherently depend on individual perceptions. Despite efforts to mitigate this through multiple annotators and Krippendorff's Alpha calculation, the low agreement on subjective metrics highlights the challenge of achieving consistent evaluations across diverse groups. Developing more objective and standardized evaluation frameworks for LLM-generated text remains a critical area for future research.

\section*{Ethical Statements}
This study deals with online experiments with a strong commitment to ethical standards in the treatment of participants.   
Prior to participation, all participants were provided with a comprehensive explanation of the study's objectives, the procedures involved, the potential risks, and their rights as study participants. Informed consent was obtained from all individual participants involved in the study. Participants were assured of their right to withdraw from the study at any point without any adverse consequences.
To protect privacy, all data collected during the study were anonymized and securely stored. Identifiable information was removed from the dataset prior to analysis to ensure confidentiality. Participants were informed that the results of the study might be published, but privacy information would remain confidential and would not be linked to any personally identifying information.
The online nature of the experiments was designed to ensure minimal risk to participants. However, appropriate measures were taken to address any technical and privacy-related concerns associated with online data collection.

\bibliography{acl_latex}

\appendix

\section{Prompts for Text Generation with Investment Stance}
\label{appendix:prompts}
Below are various prompts designed for text generation, each adopting a specific investment stance. The tasks are framed for different roles and stances:

\vspace{2mm}
\noindent \textbf{Summarization:} \textit{As a financial analyst, you are tasked with preparing a detailed summary report on a recent earnings conference call transcript. Focus on key financial metrics.} \\
    Transcripts: \{\}

\vspace{2mm}
\noindent \textbf{Overweight analysis:} \textit{As a financial analyst, you are tasked with preparing a detailed summary report on a recent earnings conference call transcript, adopting an overweight investment stance. Focus on key financial metrics.} \\
    Transcripts: \{\}

\vspace{2mm}
\noindent \textbf{Underweight analysis:} \textit{As a financial analyst, you are tasked with preparing a detailed summary report on a recent earnings conference call transcript, adopting an underweight investment stance. Focus on key financial metrics.} \\
    Transcripts: \{\}

\vspace{2mm}
\noindent \textbf{Overweight Promotion:} \textit{As a stock promoter, you are tasked with preparing a report that offers a compelling promotion on the company, recommending an overweight investment stance based on the company's recent earnings call.} \\
    Transcripts: \{\}

\vspace{2mm}
\noindent \textbf{Underweight Promotion:} \textit{As a stock promoter, you are tasked with preparing a report that offers a cautious or skeptical perspective on the company, recommending an underweight investment stance based on the company's recent earnings call.} \\
    Transcripts: \{\}

\section{Details of HTML}
\label{sec:html}
We adopt different encoders with HTML, including BERT~\cite{devlin-etal-2019-bert}, FinBERT-Tone~\cite{finbert-tone}, and FinBERT-Sentiment~\cite{prosus-finbert}, and use Adam as the optimizer with an initial learning rate of 2e-5~\citep{yang2020html}. The model is trained for 10 epochs with a batch size of 4.

\end{document}